\documentclass[pre,showpacs,twocolumn,preprintnumbers,amsmath,amssymb]{revtex4}

\usepackage{graphicx}
\usepackage{dcolumn}
\usepackage{bm}
\usepackage{parskip}
\usepackage{subfigure}
\usepackage{amssymb}
\usepackage{amsmath}
\usepackage{amssymb}
\usepackage{graphicx}

\def\be{\begin{eqnarray}}
\def\ee{\end{eqnarray}}
\def\be{\begin{equation}}
\def\ee{\end{equation}}
\begin{document}
\title{%
Crossover from Goldstone to critical fluctuations: Casimir forces in confined O${\bf(n)}$ symmetric systems
}

\author{Volker Dohm}

\affiliation{Institute for Theoretical Physics, RWTH Aachen
University, D-52056 Aachen, Germany}

\date {11 January 2013}

\begin{abstract}
We study the crossover between thermodynamic Casimir forces arising from long-range fluctuations due to Goldstone modes and those arising from critical fluctuations. Both types of forces exist in the low-temperature phase of O$(n)$ symmetric systems for $n>1$ in a $d$-dimensional  ${L_\parallel^{d-1} \times L}$ slab geometry with a finite aspect ratio $\rho = L/L_\parallel$. Our finite-size renormalization-group treatment for periodic boundary conditions describes the entire crossover from the Goldstone regime with a nonvanishing constant tail of the finite-size scaling function far below $T_c$ up to the region far above $T_c$ including the critical regime with a minimum of the scaling function slightly below $T_c$. Our analytic result for $\rho \ll 1$ agrees well with Monte Carlo data for the three-dimensional $XY$ model. A quantitative prediction is given for the crossover of systems in the Heisenberg universality class.

\end{abstract}
\maketitle

In the past two decades, substantial effort has been devoted to the study of thermodynamic Casimir forces \cite{kardar} that result from two fundamentally different sources in confined condensed matter systems: (i) from classical fluctuations with long-ranged correlations due to massless Goldstone modes \cite{zandi2004,vasilyev2009}, and (ii) from long-ranged critical fluctuations \cite{fisher78}. Both types of fluctuations exist in the low-temperature phase of O$(n)$ symmetric systems such as superfluids \cite{garcia}, superconductors \cite{wil-1}, XY magnets ($n=2$) \cite{dan-krech,vasilyev2009,hasenbusch2010}, and isotropic Heisenberg magnets $(n=3)$  \cite{dan-krech}. While successful analytic theories have been developed that separately describe such Casimir forces either (i) in the Goldstone-dominated regime deeply in the low-temperature phase \cite{zandi2004,vasilyev2009} or (ii) above bulk criticality \cite{KrDi92a,GrDi07,kastening-dohm,dohm2009}, there is a serious lack of knowledge concerning an analytic theory of the {\it crossover} between these two types of Casimir forces in the low-temperature phase. The goal of this Letter is to provide such a crossover theory for the case where the confining geometry has periodic boundary conditions. Such systems are well accessible to numerical studies \cite{dan-krech,vasilyev2009,hasenbusch2010}. We shall present analytic results for general $n$ that are in very good agreement with the existing Monte Carlo (MC) data for $n=2$ both in the Goldstone regime as well as in the critical region including the crossover between these regions. We also present a quantitative prediction for this crossover of the Casimir force in the $(n=3)$ Heisenberg universality class. The concept of our theory should be applicable also to the case of Dirichlet boundary conditions which are relevant to the crossover of Casimir forces in superfluids \cite{garcia} and superconductors  \cite{wil-1}.

A crucial ingredient of our approach for general $n$ is an appropriate choice of the geometry. We consider a finite $d$ dimensional $L_\parallel^{d-1} \times L$ slab geometry with a  finite aspect ratio $\rho=L/L_\parallel$. This
is well justified by the fact that all experiments and computer simulations were performed in slab geometries with a small but finite $\rho$ rather than $\rho=0$.
Recent Monte Carlo data for $\rho = 1/6$ \cite{vasilyev2009} and $\rho = 0.01$ \cite{hasenbusch2010} show that the $\rho$ - dependence of the Casimir force is quite weak for $\rho \ll 1$. The basic advantage of a finite-slab geometry (rather than the idealized  $\infty^{d-1} \times L$ film geometry studied earlier \cite{KrDi92a,wil-1,GrDi07,kastening-dohm,zandi2007,biswas2010}) is the absence of singularities of the free energy density at finite temperatures and the existence of a {\it discrete} mode spectrum with a dominant lowest mode that is amenable to a simultaneous analytic treatment of the low-temperature and the critical regions.

Our analytic treatment is based on the O$(n)$ symmetric isotropic $\varphi^4$ Hamiltonian
\begin{eqnarray}
\label{Hamiltonian}H &=& \int\limits_{V} d^d x
\big[\frac{r_0} {2} \varphi^2 +  \frac{1} {2} (\nabla
\varphi)^2 +
 u_0 (\varphi^2)^2   \big]
\end{eqnarray}
where $\varphi({\bf x})= V^{-1}\Sigma_{{\bf k}}\varphi_{{\bf k}} e^{i{\bf k}{\bf x}}$ is an $n$-component field  in a finite volume $V=L_\parallel^{d-1}L $  with periodic boundary conditions. The summation $\Sigma_{{\bf k}}$ runs  over discrete ${\bf k}$ vectors including ${\bf k}={\bf 0}$ up to some cut off $\Lambda$. The fundamental quantity from which the Casimir force per unit area $F_{{\text Cas}}=-\partial [Lf^{{\text ex}}]/\partial L$
can be derived is the excess free energy density (divided by $k_BT$) $f^{{\text ex}}=f-f_b$ where
\be
f(T,L,L_\parallel)= -V^{-1}\ln \int {\cal D} \varphi \exp(-H)
\ee
and  $f_b \equiv \lim_{V\rightarrow \infty}f$ are the free energy densities of the finite system and the bulk system, respectively.

It is expected that, for isotropic systems near criticality and for large $L$ and $L_\parallel$,  $F_{{\text Cas}}$ can be written in a finite-size scaling form \cite{pri}
\begin{equation}
\label{3}
F_{{\text Cas}}(t, L, L_\parallel)=L^{-d} X(\tilde x,\rho)
\end{equation}
with the scaling variable $ \tilde x=t(L/\xi_{0+})^{1/\nu}$, $t=(T-T_c)/T_c$ where $\xi_{0+}$ is the amplitude of the bulk correlation length  above $T_c$. As noted for the case of film geometry ($\rho \to 0$) \cite{vasilyev2009}, the scaling form (\ref{3}) applies to the low-temperature region where the scaling function saturates at a nonzero negative value $X( -\infty, 0) < 0$ for $n>1$ \cite{zandi2004,vasilyev2009,biswas2010}. So far, however, no analytic calculation of the function $X(\tilde x,0)$ for systems with $n>1$  and  periodic boundary conditions has been performed that describes $X$ in the whole low-temperature region $-\infty \leq \tilde x \leq 0$ \cite{footnote1}. It is the goal of this paper to describe the full crossover in terms of a single scaling function $X(\tilde x,\rho)$, for general $n>1$ and  small $\rho$, from the Goldstone-dominated behavior for  $\tilde x \to -\infty$ up to the high-temperature behavior for  $\tilde x \gg 1$ including the critical region $0 \leq |\tilde x| \sim O(1)$ above and below $T_c$.

An important conceptual difference between the previous perturbation approach within the $\varphi^4$ theory for $(n=1)$ systems with a discrete (Ising-like) symmetry \cite{dohm2009} and  for systems with a continuous symmetry $(n>1)$ presented in this paper is the following. For $n = 1$, two separate perturbation approaches were necessary for describing finite-size effects both in the central finite-size regime near criticality and those in the low-temperature phase far below $T_c$ in order to capture the two-fold (spin up and spin down) degeneracy characteristic of the ground state of Ising-like systems. For $n > 1$, there exist both longitudinal and transverse fluctuations. The latter correspond to orientational changes that permit the order parameter fluctuations to exhaust the full phase space since there exists no large barrier like that between the purely longitudinal spin-up and spin down configurations of large Ising-like systems far below $T_c$. As a consequence, a single perturbation ansatz suffices for $n>1$ to capture both the critical fluctuations and the fluctuations due to the Goldstone modes. This results in a smooth description of the crossover between the two different regimes without the necessity of matching two separate pieces of the theory.

We decompose $\varphi = \Phi + \sigma $ into a homogeneous lowest-mode
$\Phi=V^{-1}\int d^d x \varphi({\bf x})$ and  higher-mode fluctuations $\sigma({\bf x})= V^{-1}\Sigma_{{\bf k}\neq {\bf 0}}\varphi_{\bf k} e^{i{\bf k}{\bf x}}$. We further decompose ${\bf\sigma}({\bf x})$  into "longitudinal" and "transverse" parts ${\bf\sigma}({\bf x}) = {\bf\sigma}_{\rm L}({\bf x}) + {\bf\sigma} _{\rm T}({\bf x})$ which are parallel and perpendicular with respect to $\Phi$. Correspondingly, the Hamiltonian $H$ is decomposed as
$H = H_0(\Phi) + \widetilde H (\Phi, \sigma)$ where
\begin{eqnarray}
\label{lowHamilton}
H_0 (\Phi) = V
\left[\frac{1}{2} r_0 \Phi^2 + u_0 {(\Phi^2)}^2  \right],
\end{eqnarray}
\begin{eqnarray}
\label{higher Hamilton}
\widetilde H (\Phi, \sigma)= \int\limits_{V} d^d x\big\{
\frac{1} {2} \big[         r_{0{\rm L}} {{\bf\sigma}_{\rm L}}^2 +  r_{0{\rm T}} {{\bf\sigma}_{\rm T}}^2 + (\nabla
{\bf\sigma}_{\rm L})^2  \nonumber \\+ (\nabla
{\bf\sigma}_{\rm T})^2 \big] +  4 u_0  \Phi\sigma_{\rm L} \sigma^2 + u_0 (\sigma^2)^2 \big\}
\end{eqnarray}
with the longitudinal and transverse parameters
$r_{0{\rm L}}(\Phi^2) = r_0+12u_0 \Phi^2, r_{0{\rm T}}(\Phi^2) = r_0+4u_0\Phi^2$. The free energy density $f$ is then calculated by first integrating over $\sigma$ and subsequently over $\Phi$.

An important reference quantity of our theory is the $n$-dependent lowest-mode average $M_0^2=\int d^n  \Phi  \Phi^2\exp \left[-H_0(\Phi)\right]/\int d^n  \Phi \exp \left[-H_0(\Phi)\right]$. The physical importance of the quantity $M_0^2$ is related to the fact that the main contribution of the integration over $ \Phi$ comes from the region around $\Phi^2\approx M_0^2$ and that this quantity is relevant in the whole range $-\infty < r_0 < \infty$ above and below $T_c$. This provides the justification for replacing the $\Phi^2$-dependent parameters $r_{0{\rm L}}(\Phi^2)$  and $r_{0{\rm T}}(\Phi^2)$ in $\widetilde H$ by $\bar r_{0{\rm L}}\equiv r_{0{\rm L}}(M_0^2)$ and $\bar r_{0{\rm T}}\equiv r_{0{\rm T}}(M_0^2)$. The second approximation is to neglect the effect of the higher-mode parts $\propto \Phi\sigma_{\rm L} \sigma^2$ and $\propto (\sigma^2)^2$ on the finite-size properties (but not on the bulk critical exponents which will be incorporated via Borel-resummed field-theoretic functions). No further approximation (such as an $\varepsilon =4-d$ expansion) will be made in the subsequent renormalization-group treatment at fixed $d$. Our approach goes beyond a naive Gaussian approximation not only because of the fourth-order term $\sim u_0{( \Phi^2)}^2$ in $H_0$ but also
because of the {\it size-dependent} couplings $ \sim u_0 M_0^2 {\sigma_{\rm L}}^2$ and
$\sim u_0 {M_0}^2 {\sigma_{\rm T}}^2$ between the lowest-mode average and the higher modes which arise from the terms $ \bar r_{0{\rm L}} {\sigma_{\rm L}}^2 $  and $\bar r_{0{\rm T}} {\sigma_{\rm T}}^2$ , respectively. After integration over $\sigma$, we obtain the unrenormalized free energy density
\begin{eqnarray}
\label{bare-free}
f = f_0 - \frac{1}{V}\ln \Big\{ \int d^n \Phi \exp \left[-H_0(\Phi)\right]\Big\}\nonumber \\ + \frac{1}{2}S_0(\bar r_{0\rm L}) + \frac{n-1}{2}S_0(\bar r_{0\rm T}),
\end{eqnarray}
with $S_0(r)=V^{-1} {\sum_{\bf k\neq0}} \ln ( r + \mathbf k^2)$
where $f_0$ is independent of $r_0$ and $u_0$.

For finite $V$, the transverse parameter  $\bar r_{0{\rm T}}$ remains positive in the whole range $-\infty < r_0 < \infty$ and interpolates smoothly between the large-volume limit above and below $T_c$,
\be
\label{rprimetrans} \lim_{V \rightarrow \infty} \bar r_{0\rm T} = \left\{
\begin{array}{r@{\quad\quad}l}
                 r_0 & \mbox{for} \;\;\; r_0\geq0\;, \\  0 & \mbox{for} \;\;\;
                 r_0\leq0\;.
                \end{array} \right.
\ee
The vanishing of the transverse parameter $\bar r_{0\rm T}$ is the characteristics of the massless Goldstone modes which are the origin of long-range correlations and the Casimir force well below $T_c$. In higher-order perturbation theory for the bulk system, spurious (infrared) singularities arise due to the vanishing of $\bar r_{0\rm T}$  \cite{str}. Within our approximation, such spurious singularities do not yet appear since $S_0(\bar r_{0\rm T})$ has a finite large-volume limit below $T_c$.

The bare expression (\ref{bare-free}) does, of course, not yet correctly describe the crossover from the Goldstone to the critical regime. Both additive and multiplicative renormalizations are necessary, after  subtracting  a nonsingular bulk part $f_{ns}$. Integration of the renormalization-group equation then leads to the scaling form $f_s(t, L, L_{\parallel})=L^{-d} F(\tilde x, \rho)$
for the singular part $f_s = f - f_{ns}$. We have performed these steps within the minimal renormalization scheme at fixed dimensions $2<d<4$ \cite{dohm1985}. The result reads
\begin{eqnarray}
\label{free-energy-scaling} &&F(\tilde x, \rho)= -\; A_d \;\Bigg\{
\frac{\tilde l^d}{4d} \; + \;\frac{\nu\;{Q^*}^2 \tilde x^2 \tilde
l^{- \alpha/\nu}}{2\alpha} \;B(u^*)\nonumber\\ && - \frac{(n-1)}{\varepsilon}\Bigg[\frac{l_{\rm T}^2}{4 \tilde l^\varepsilon }-\frac{l_{\rm T}^{d/2}}{d}\Bigg]\Bigg\} \nonumber\\ && + \; \rho^{d-1} \Bigg\{  -\frac{n}{2} \ln\Big(\frac{2\pi^2A_d^{1/2}}{\tilde l^{\varepsilon/2}\rho^{(d-1)/2}[\Gamma(n/2)]^{2/n}{u^*}^{1/2}}\Big)\nonumber\\&& -\;  \ln \Big(2 \int\limits_0^\infty d s s^{n-1}\;\exp \big[-
\frac{1}{2}  \tilde y(\tilde x, \rho) s^2 - s^4\big]\Big) \nonumber\\
&&+ \frac{1}{2}{\cal J}_0( {\tilde l}^2, \rho)+  \frac{n-1}{2}{\cal J}_0( l_{\rm T}, \rho)\Bigg\},
\end{eqnarray}
\begin{eqnarray}
\label{calJ3}
&&{\cal J}_0(z,\rho )= \rho^{1-d}  \int\limits_0^\infty
dy y^{-1}
  \Bigg\{\exp {\left[-z y/(4\pi^2)\right]}
  \nonumber\\ &&\times \left[
  (\pi / y)^{d/2}
  - \;\Big[\rho K(\rho^2y)\Big]^{d-1}K(y) + \rho^{d-1} \right] \nonumber\\ &&   - \rho^{d-1}\exp (-y)\Bigg\}
\end{eqnarray}
with $ K(y) = \sum^{\infty}_{m= -\infty} \;\exp (- y m^2)$ and with
\begin{eqnarray}
\label{flowrtransasymp}
l_{\rm T}(\tilde x,\rho)= \tilde l^2 -8{\tilde
l}^{\varepsilon/2}{u^*}^{1/2}\rho^{(d-1)/2}A_d^{-1/2}\vartheta_2 (\tilde y) ,
\end{eqnarray}
\be
\vartheta_2 (y) =\int\limits_0^\infty d s \;s^{n+1}e^{-
\frac{1}{2} y s^2 - s^4}/\int\limits_0^\infty d s \;s^{n-1}e^{-
\frac{1}{2} y s^2 - s^4},
\ee
where  $\tilde l (\tilde x,\rho)$ and   $\tilde
y(\tilde x,\rho)$ are determined implicitly by
\begin{subequations}
\label{y}
\begin{align}
\label{y-a}
\tilde y + 12 \vartheta_2(\tilde y) = \rho^{(1-d)/2}
\tilde l^{d/2} A_d^{1/2} {u^*}^{-1/2},
\\
\label{y-b}
\tilde y =\; \tilde x\;Q^*\;\tilde l^{- \alpha / (2
\nu)}\rho^{(1-d)/2} A_d^{1/2} {u^*}^{-1/2}.
\end{align}
\end{subequations}
The fixed point value of the renormalized four-point coupling $u$ is denoted by $u^*$.
The quantities $ Q^* =Q(1,u^*,d) $ and $B(u^*)$ are the fixed point values of the $n$ dependent amplitude function $Q (1, u, d)$ of the second-moment bulk correlation length above $T_c$ and of the field-theoretic function $B(u)$ related to the additive renormalization of bulk theory \cite{dohm1985}, respectively. $A_d = \Gamma(3-d/2)[2^{d-2} \pi^{d/2}(d-2)]^{-1}$ is an appropriate geometric factor.

\begin{figure}[!ht]
\begin{center}
\subfigure{\includegraphics[clip,width=7.92cm]{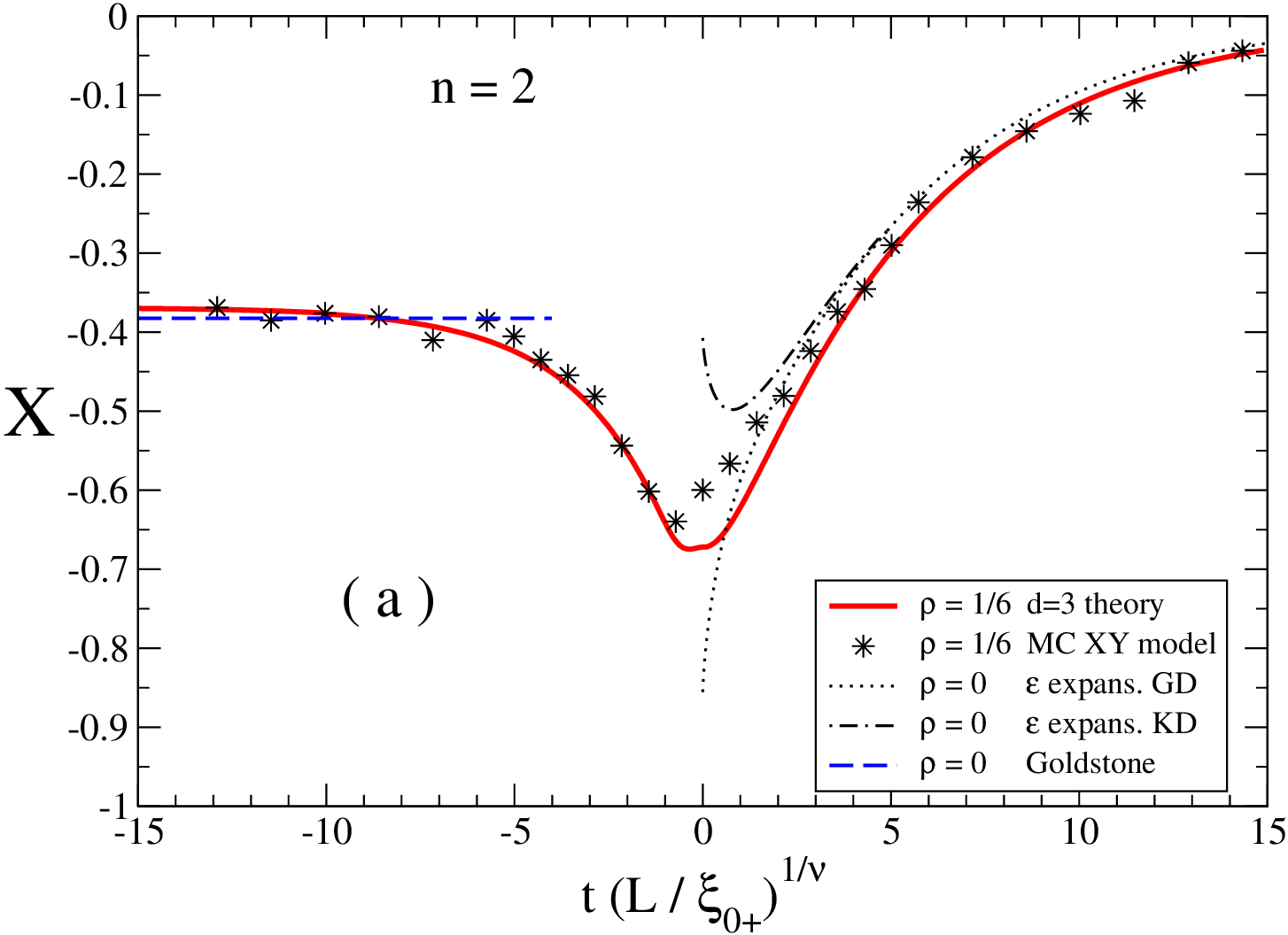}}
\subfigure{\includegraphics[clip,width=7.92cm]{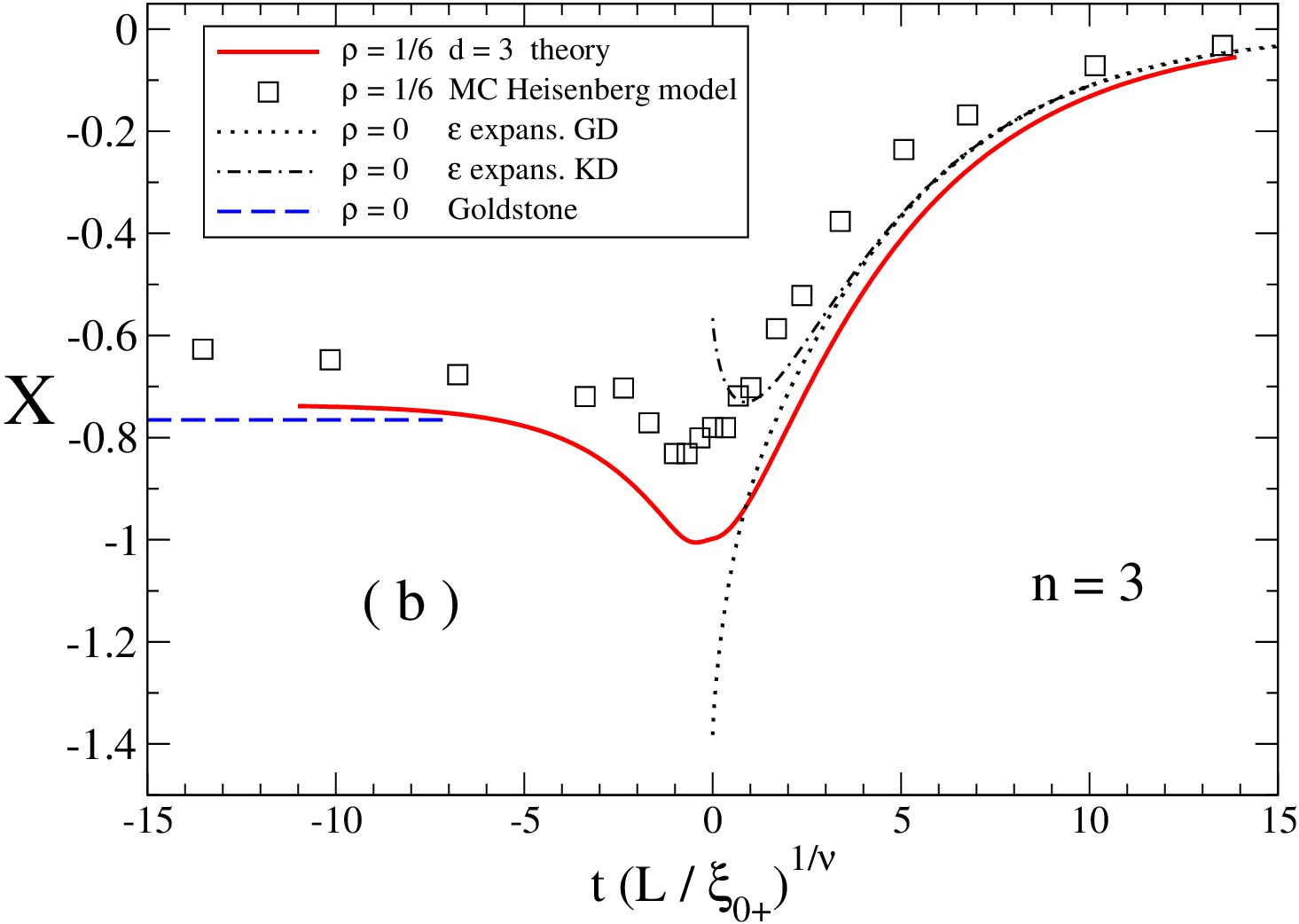}}
\end{center}
\caption{(Color online) Scaling function $X(\tilde x, \rho)$ of the Casimir force as a function of $\tilde x = t (L/\xi_{0+})^{1/\nu}$   (a) for $n=2$ and (b) for $n=3$ in three dimensions. Solid lines: $d=3$ theory for $\rho=1/6$ as calculated from (\ref{free-energy-scaling}),(\ref{casimir-scalingfunction}). Dot-dashed and dotted lines: $\varepsilon$ expansion results for $\rho=0$  \cite{KrDi92a,GrDi07}. Horizontal dashed lines: Casimir amplitude $X(-\infty,0)=-(n-1) \zeta(3)/\pi$.  MC data (a) from \cite{vasilyev2009} for  the XY model and (b) from  \cite{dan-krech} for  the  Heisenberg model.}
\end{figure}

Eq. (\ref{free-energy-scaling}) is valid for general $n$ and $2<d<4$  above, at, and below $T_c$ including the Goldstone regime for $n>1$.  It incorporates the correct bulk critical exponents $\alpha$ and $\nu$ and the complete bulk function $B(u^*)$ (not only in one-loop
order). $F(\tilde x, \rho)$  is an analytic function of  $\tilde x$ at finite $\rho$, in agreement with general analyticity requirements. For $n=1$, the function (\ref{free-energy-scaling}) is not applicable to the region well below $T_c$ as it does not capture the two-fold degeneracy of the ground state of Ising-like systems, as discussed in \cite{dohm2009}.

To obtain the scaling function $X$ of the Casimir force we consider the singular part of the bulk free energy density $f^\pm_{s,b}(t)=A^\pm |t|^{d\nu}$. It can be written as $f^\pm_{s,b}= L^{-d}{F}^\pm_b(\tilde x)$ where ${F}^\pm_b(\tilde x)$ is the bulk part of $F(\tilde x, \rho)$. It is derived from (\ref{free-energy-scaling}) in the limit of large $|\tilde x|$. The scaling function of $f^{{\text ex}}$ is then given by $F^{{\text ex}}(x,\rho)= F(x,\rho) - {F}^\pm_b(x)$ from which we calculate the desired scaling function
\begin{eqnarray}
\label{casimir-scalingfunction} X(\tilde{x},\rho) =(d-1)  F^{ex} (\tilde{x},\rho) -
\frac{\tilde{x}}{\nu}
\frac{\partial F^{ex}(\tilde{x},\rho)}{\partial\tilde{x}}- \rho
\frac{\partial F^{ex}(\tilde{x},\rho)}{\partial \rho}.
\end{eqnarray}
In the limit $\rho \to 0, \tilde x \to -\infty$, our function $X(\tilde{x},\rho)$ yields the finite low-temperature amplitude
\begin{align}
\label{casimir-film}
X(-\infty,0)=-(n-1)(d-1)\pi^{-d/2}\Gamma(d/2)\zeta(d)
\end{align}
of the Goldstone regime in film geometry. For $n=2,d=3$ this agrees with Eq. (25) of \cite{vasilyev2009}.

MC data are available for both the three-dimensional XY  \cite{vasilyev2009} and Heisenberg \cite{dan-krech} models with the aspect ratio $\rho=1/6$. The comparison of our result for $\rho=1/6$ with the MC data is shown in Figs. 1 (a) and (b) for $n=2$ and $n=3$, respectively. Here we have employed the following numerical values \cite{CDS1996,larin}: $\nu= 0.671, 0.705$, $\alpha =  -0.013, -0.115$, $u^*=  0.0362, 0.0327$, $Q^*=  0.939, 0.937$, $B(u^*) = 1.005, 1.508$, for $n=2,3$, respectively. Also shown are the $\varepsilon$ expansion results for $\rho=0$ from \cite{KrDi92a,GrDi07} for $T \geq T_c$ (dot-dashed and dotted lines). The horizontal dashed lines represent the $\rho=0$ Casimir amplitude (\ref{casimir-film}) due to the Goldstone modes. For $n=2$ there is an excellent overall agreement of our result (solid line) with the MC data \cite{vasilyev2009} including the crossover from the Goldstone-dominated region to the critical region.

For $n=3$ [Fig. 1 (b)] there are systematic deviations between all theoretical curves and the MC data of \cite{dan-krech}. A similar discrepancy with the MC data of \cite{dan-krech} exists also for $n=2$ [not shown in our Fig. 1 (a)] as shown in Fig. 6 of \cite{vasilyev2009} for the XY model. As noted in \cite{vasilyev2009}, these discrepancies might be due to the uncertainty in the normalization factor used for the MC data in \cite{dan-krech}. More accurate MC simulations are desirable in order to test our prediction for $n=3$.

\begin{figure}[!h]
\includegraphics[width=80mm]
{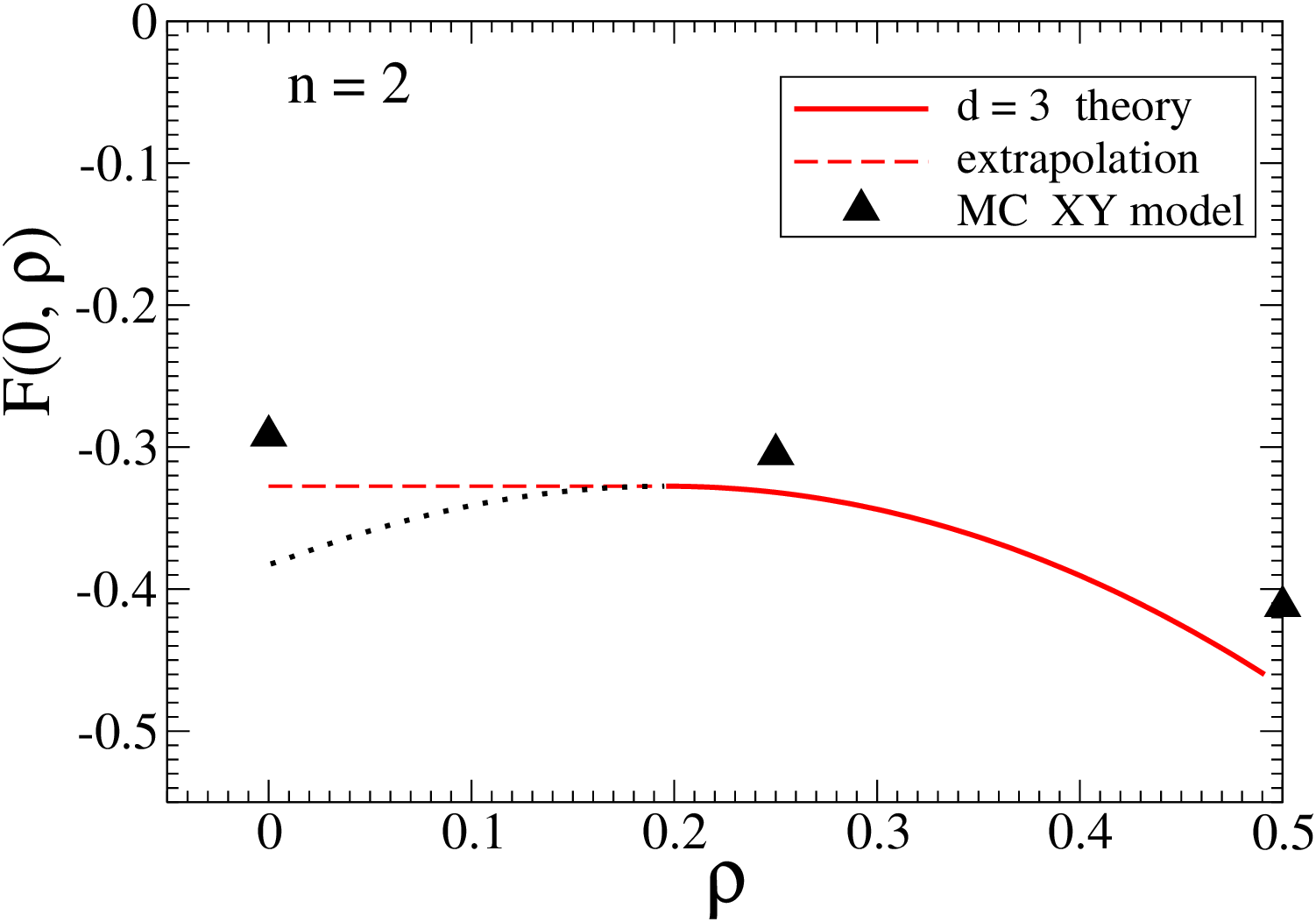} \caption{(Color online) Critical amplitude   $F(0, \rho)$, (\ref{free-energy-scaling}), for $n=2$ and $d=3$ at $T_c$ (solid line). The maximum $-0.3275$ is at $\rho_{{\text max}}= 0.196$. The dashed line is the extrapolation from $\rho = \rho_{{\text max}}$ to $\rho =0$ (film geometry). The dotted line represents (\ref{free-energy-scaling}) at $T_c$ for $\rho < \rho_{{\text max}}$. MC estimate (triangles) for the $d=3$ XY model   for $\rho = 0.01,1/4,1/2$ from \cite{hasenbusch2011} .}
\end{figure}

Finally we discuss the question as to the expected range of applicability of our theory at finite $\rho$. For this purpose we consider the finite-size amplitude $F(0, \rho)$ of Eq. (\ref{free-energy-scaling}) at $T=T_c$. We expect a monotonic $\rho$ - dependence of $F(0, \rho)$ on the basis of the monotonicity hypothesis inferred from previous results for $n=1$ \cite{dohm2009,hucht2011} and $n=\infty$ \cite{dohm2009}. A comparison with MC data for the three-dimensional XY model \cite{hasenbusch2011} is shown in Fig. 2 in the range $0\leq \rho \leq 1/2$. The MC data show a very weak $\rho$ - dependence for $\rho \ll 1$. Our result is in reasonable agreement with the MC data, except that the monotonicity of the $\rho$ - dependence is not reproduced by our theory for small $\rho < \rho_{{\text max}}=0.196$ (dotted portion of the curve). The deterioration of the quality of our theory for $\rho \to 0$ is to be expected since the separation between the lowest mode and the higher modes goes to zero in this limit. For this reason it is expected that Eqs. (\ref{free-energy-scaling}) - (\ref{y}) are not quantitatively reliable for $\rho \ll 0.2$ in some region $|\tilde x|\ll1$ close to $T_c$. In particular, the scaling function of the film free energy density $F_{{\text film}}(\tilde x)= \lim_{\rho \to 0} F(\tilde x,\rho)$ obtained from (\ref{free-energy-scaling}) is not expected to be reliable for $|\tilde x| \ll 1$. This function has, in fact, an artificial cusp-like singularity at $\tilde x=0$ similar to that of previous approximate $\rho=0$ theories  \cite{KrDi92a,wil-1,GrDi07,kastening-dohm,zandi2007,biswas2010}. Our monotonicity criterion provides a theory-internal argument against the reliability of approximate $\rho=0$ results in the region $| \tilde x| \ll 1$.
Nevertheless, assuming a negligible $\rho$ - dependence for  $\rho < \rho_{{\text max}}$ we obtain from the extrapolation of the maximum in Fig. 2 to $\rho=0$ (dashed line in Fig. 2) our prediction
\be
F_{{\text film}}(0)\approx F(0, \rho_{{\text max}})=-0.3275
\ee
for film geometry at bulk $T_c$.
This is in acceptable agreement with the MC estimate $-0.292$ at $\rho=0.01$ \cite{hasenbusch2011} shown in Fig. 2.

It would be interesting and important to extend our approach to the most relevant case of Dirichlet boundary conditions in order to explain the experimental results for the Casimir force in superfluid $^4 {\text He}$ \cite{garcia} for which there exists no satisfactory analytic theory so far.

The author is grateful to D. Dantchev, M. Hasenbusch, and O. Vasilyev for providing  the MC data of \cite{dan-krech,hasenbusch2011,vasilyev2009} in numerical form.

\end{document}